\documentclass{svjour2}
\usepackage{graphicx}
\usepackage{amsbsy}

\journalname{Celestial Mech Dyn Ast}
\begin{document}

\title{Stability criteria for hierarchical triple systems}
\titlerunning{Stability criteria}
\author{Nikolaos Georgakarakos}
\authorrunning{Nikolaos Georgakarakos}
\institute{Institute for Materials and Processes, School of
Engineering and Electronics, The University of Edinburgh, Mayfield Road, Edinburgh EH9 3JL, UK\\
\email{georgakarakos@hotmail.com}}

\date{Received: date / Accepted: date}

\maketitle

\begin{abstract}
In this paper, we  give a summary of stability criteria that have been derived
for hierarchical triple systems over the past few decades.
We give a brief description and we discuss the criteria that are based on
the generalisation of the concept of zero velocity surfaces of the restricted
three body problem, to the general case.  We also present criteria that have to do
with escape of one of the bodies.  Then, we talk about the criteria that have been
derived using
data from numerical integrations.  Finally, we report on criteria that involve the concept
of chaos.
In all cases, wherever possible, we discuss advantages
and disadvantages of the criteria and the methods their derivation was based on, and
some comparison is made in several cases.

\keywords{Celestial mechanics, Three body problem, Stability}
\end{abstract}

\section{INTRODUCTION}
 The three body problem is one of the most fascinating topics in
mathematics and celestial mechanics.  The basic definition of the
problem is as follows: three point masses (or bodies of spherical
symmetry) move in space, under their mutual gravitational attraction;
 given their initial conditions, we want to determine their
subsequent motion.

Like many mathematical problems, it is not as simple as it sounds.
Although the two body problem can be solved in closed form by
means of elementary functions and hence we can predict the
quantitative and qualitative behaviour of the system, the three body
problem is a complicated nonlinear problem and no similar type of
solution exists.  More precisely, the former is integrable but the
latter is not (if a system with n degrees of freedom has n
independent first integrals in involution, then it is integrable;
that is not the case for the three body problem).

One issue that is of great interest in the three body problem, is
the stability (and instability) of triple systems.
The stability (and instability) of  triple systems is
an intriguing problem which remains unsolved up to date.
It has been a subject of study by many people, not only
because of the intellectual challenge that
poses, but also because of its importance in many areas
of astronomy and astrophysics, e.g. planetary and star
cluster dynamics.

In this work, we review the three body stability criteria
that have been derived over the past few decades.
We deal with the gravitational non-relativistic three-body problem and
we concentrate on hierarchical triple systems.  By hierarchical, we mean
systems in which we can distinguish two different motions: two of the bodies
form a binary and move around their centre of mass, while the third body
is on a wider orbit with respect to the binary barycentre.  This may not
be the most strict definition of a hierarchical triple system (e.g. see Eggleton
and Kiseleva 1995), but we use that one in order to cover as many triple system
configurations as possible.

We would also like to point out that some of the criteria may apply to systems that
are not hierarchical or
they are marginally hierarchical (e.g. Wisdom's criterion for resonance overlap),
according to the definition given in the previous paragraph.
However, as they are related to other criteria that refer to hierarchical systems, we
felt that we should mention them too.

\section{STABILITY CRITERIA}

There are two main types of stability criteria, depending on how they
were derived: analytical and numerical.  Following that
classification, we are going to present the analytical criteria
first and then we will discuss the criteria that have been derived
from numerical integrations.  Finally, we present criteria that are
based on the concept of chaos.

Throughout the next paragraphs, we decided that it would be better
 if we kept the notation that each author used (with a few exceptions
for the benefit of the reader).

\subsection{Analytical Criteria}

The derivation of analytical stability criteria in the three body
problem has been
dominated by the generalisation of the concept of surfaces
of zero velocity of the restricted three-body problem, first
introduced by Hill (1878a,1878b,1878c).  It is known that in the circular
restricted three body problem, there are regions in physical
space where motion can and cannot occur.  These regions are
determined by means of the only known integral of the
circular restricted problem, the so called Jacobi constant.
This notion has been extended to the general three body problems
 by several authors: Golubev (1967, 1968a, 1968b), Saari (1974),
who used
an inequality similar to Sundman's, Marchal and Saari (1975),
who used Sundman's inequality, Bozis (1976), who used algebraic
manipulations of the integrals of motion in the planar three
 body problem, Zare (1976, 1977), who made use of Hamiltonian dynamics;
Saari (1984, 1987), who produced 'the best possible configurational
 velocity surfaces'.  Also, Sergysels (1986), derived zero velocity surfaces
for the general three dimensional three body problem, by using the
method of Bozis (1976) and a rotating frame that does not take
into account entirely the rotation of the three body system.
Finally, Ge and Leng (1992) produced the same result as Saari
(1987), using a modified version of the transformation given in
Zare (1976). Easton (1971), Tung (1974) and Mialni and Nobili
(1983) also discussed the topology of the restrictive surfaces.

The quantity ${c^{2}H}$, where
${c}$ is the angular momentum and ${H}$ is the energy of the three body system,
controls the topology of the restrictive surfaces and it is the analog of the
Jacobi constant of the circular restricted problem.

Szebehely (1977) and Szebehely and Zare (1977), using two body
approximations, produced an expression for ${c^{2}H}$, which
involved the masses, the semi-major axes and the eccentricities of
the system. Then,  that expression was compared with the critical
value  ${(c^{2}H)_{crit}}$ at the collinear Lagrangian points,
which determine the openings and closings of the zero velocity
surfaces.  If the value of ${c^{2}H}$ for a given triple
configuration was smaller than the one at the inner Lagrangian
point, then there could be no exchange of bodies, i.e the system
was Hill stable. Although there was some discussion on the effect
of the inclination, the derivation was for coplanar orbits.

Marchal and his collaborators (Marchal and Saari 1975, Marchal and Bozis 1982),
produced a generalisation of the Hill curves to the general three dimensional three body problem
by using the quantity ${\rho / \nu}$ as the controlling parameter of the restrictive surfaces,
where ${\rho}$ is the mean quadratic distance, ${\nu}$ is the mean harmonic distance and they are
defined by the following equations:
\begin{equation}
M^{*}\rho^{2}=m_{1}m_{2}r^{2}_{12}+m_{1}m_{3}r^{2}_{13}+m_{2}m_{3}r^{2}_{23}
\end{equation}
\begin{equation}
\frac{M^{*}}{\nu}=\frac{m_{1}m_{2}}{r_{12}}+\frac{m_{1}m_{3}}{r_{13}}+\frac{m_{2}m_{3}}{r_{23}},
\end{equation}
where ${M^{*}=m_{1}m_{2}+m_{1}m_{3}+m_{2}m_{3}}$ and ${r_{ij}}$ is the distance between
${m_{i}}$ and ${m_{j}}$.

Walker et al. (1980) derived the critical surfaces in terms of the parameters
\begin{displaymath}
\epsilon^{23}=\frac{m_{1}m_{2}}{(m_{1}+m_{2})^{2}}\alpha^{2}_{23}
\hspace{0.5cm}\mbox{and}\hspace{0.5cm}
\epsilon_{32}=\frac{m_{3}}{m_{1}+m_{2}}\alpha^{3}_{23}
\end{displaymath}
with ${\alpha_{23}=\rho_{2}/\rho_{3}}$ (${\rho_{2}}$ is the
distance between ${m_{1}}$ and ${m_{2}}$, ${\rho_{3}}$ is the
distance between the centre of mass of ${m_{1}}$ and ${m_{2}}$,
and ${m_{3}}$). ${\epsilon^{23}}$ measures the disturbance of
${m_{3}}$ by the binary, while ${\epsilon_{32}}$ is a measure of
the disturbance of the binary by ${m_{3}}$.  Thus, for a given
triple configuration, they  evaluated the ${\epsilon}$ quantities
and determined whether the system was Hill stable or not.

Walker and Roy (1981) investigated the effect that the
eccentricities had on the stability limit, as the Walker et al.
(1980) derivation applied only for coplanar, initially circular
and corotational triple systems. They paid particular attention to
the initial orbital phases of the system and they found that the
critical value of ${\alpha=\alpha_{cr}}$ (${\alpha}$ being the
semi-major axis ratio of the two orbits) could be affected by up
to ${20\%}$. Similar work was also done in Valsecchi et al.
(1984), but instead of using two body expressions for the angular
momentum and energy of the system as Walker et al. (1980) did,
they used the exact expressions; however the disagreement between
the two methods was very small.   This was also confirmed by
Kiseleva et al. (1994b), who used the exact expressions for the
angular momentum and energy to evaluate ${X_{SZ}}$ (the critical
initial semi-major axis for the Szebehely-Zare criterion).  They
found that their value was always larger by at most ${5\%}$
compared to the one obtained by two body approximations.

Roy et al. (1984) computed the distance of the closest approach of
${m_{2}}$ to ${m_{3}}$ for a coplanar, corotational, hierarchical
three body system (with ${m_{2}<m_{1}}$ for the inner binary) and
derived a condition for stability by manipulating the
angular momentum and energy integrals. They ended up with the
following inequality:

\begin{equation}
\label{roy1984}
-2s \leq [\mu\frac{(1-k)^{2}}{1-\mu}+\frac{\mu_{3}}{1+\mu_{3}}][\mu\frac{(1-\mu)^{2}}{1-k}+\frac{\mu\mu_{3}}{k}+
\frac{\mu_{3}(1-\mu)^{2}}{1-k\mu}]^{2},
\end{equation}
where
\begin{displaymath}
s=\frac{c^{2}H}{(m_{1}+m_{2})^{5}}, \hspace{0.3cm}\mu=\frac{m_{2}}{m_{1}+m_{2}},
\hspace{0.3cm}\mu_{3}=\frac{m_{3}}{m_{1}+m_{2}},
\end{displaymath}
${k}$ is defined by the relation ${\rho_{2}(1-\mu)=\rho_{3}(1-k)}$
(${\rho_{2}}$ and ${\rho_{3}}$ are the magnitudes of the two
Jacobian vectors of the hierarchical triple system) and it
represents the distance of closest approach of ${m_{2}}$ to
${m_{3}}$. If there exists a dynamical barrier between ${m_{2}}$
and ${m_{3}}$, then, there will be values of ${k}$ for which
inequality (\ref{roy1984}) will not be satisfied. The largest of
these values will give the measure of the closest approach of the
two orbits.  Their result was in agreement with the ${c^{2}H}$
criterion.

The concept of Hill type surfaces that pose restrictions to the
motion of three body systems, has also been used to study the
motion in special cases.

Szebehely (1978), in the context of the circular restricted three body problem, derived a simple condition for a
satellite to remain in orbit around the smaller primary in presence of the perturbations of the larger one. The condition
is:
\begin{equation}
(\rho_{2})_{max} \leq (\frac{\mu}{81})^{\frac{1}{3}},
\end{equation}
${\rho_{2}}$ being the radius of the satellite circular motion
around its primary ${m_{2}}$ and ${\mu =m_{2}/(m_{1}+m_{2})}$. The above
condition is valid for both prograde and retrograde motion.

Markellos and Roy (1981) obtained a more accurate result for the same problem:
\begin{equation}
R^{D}_{max}=1.4803(\frac{\mu}{81})^{\frac{1}{3}}[1-1.73(\frac{\mu}{81})^{\frac{1}{3}})]+O(\mu)
\end{equation}
for prograde orbits and
\begin{equation}
R^{D}_{max}=0.8428(\frac{\mu}{81})^{\frac{1}{3}}[1-0.55(\frac{\mu}{81})^{\frac{1}{3}})]+O(\mu)
\end{equation}
for retrograde orbits, where ${R^{D}_{max}}$ corresponds to
${(\rho_{2})_{max}}$ of Szebehely (1978) and again, ${\mu =m_{2}/(m_{1}+m_{2})}$.

Walker (1983) investigated the Hill-type stability
of a coplanar, with initially circular orbits, hierarchical three body system, where the total mass of the binary
was small compared to the mass of the external body (
e.g. satellite-planet-star).  His results were in good agreement with Szebehely (1978) and Markellos and Roy (1981).

Donnison and Williams (1983, 1985) used the ${c^{2}H}$ condition to determine the Hill stability of coplanar hierarchical
three body
systems with ${m_{1} \gg m_{2}, m_{3}}$ (${m_{1}}$ and ${m_{2}}$ form the inner binary).
Using two body approximations for the angular momentum and the energy
of the system
and taking advantage of the fact that one of the masses was much greater that the other two, they concluded that
their system
was stable (in terms of exchange) when the following condition was satisfied:
\begin{equation}
e^{2}_{max}\leq \frac{\lambda (\epsilon_{1}-3)+\lambda^{2}(\epsilon_{2}-3)}{1+\lambda\epsilon_{1}+\lambda^{2}\epsilon_{2}+
\lambda^{3}},
\end{equation}
where
\begin{displaymath}
\epsilon_{1}=(\frac{a_{1}}{a_{2}}) \pm 2(\frac{a_{2}}{a_{1}})^{\frac{1}{2}},
\epsilon_{2}=(\frac{a_{2}}{a_{1}}) \pm 2(\frac{a_{1}}{a_{2}})^{\frac{1}{2}},
\lambda=\frac{m_{3}}{m_{2}},
\end{displaymath}
${a_{1}}$ and ${a_{2}}$ are the semi-major axes of the inner and outer orbit respectively; the plus sign corresponds to
prograde motion, while the minus sign to retrograde motion.Finally, ${e_{max}}$ is the largest of either inner or
outer eccentricity.

Donnison (1988), using the same approach mentioned above, investigated the stability of low mass binary systems
moving on elliptical orbits in the presence of a large third mass, i.e. ${m_{3}\gg m_{1}+m_{2}}$.

Brasser (2002)
dealt with systems where ${m_{2}}$ was smaller that the other two masses, which were of comparable size
(${m_{1}}$ and ${m_{2}}$ form the inner binary).

Gladman (1993), based on the work done by Marchal and Bozis (1982), produced analytical formulae for the critical
separation ${\Delta_{c}}$ that two planets ${m_{1}}$ and ${m_{2}}$, orbiting a star ${m_{3}}$, should have
in order to be Hill stable.  He derived the following formulae (to lowest order):

\noindent (i) for initially circular orbits
\begin{equation}
\Delta_{c}\approx 2.40(\mu_{1}+\mu_{2})^{\frac{1}{3}}
\end{equation}
(ii) equal mass planets, small eccentricities ${(\mu_{1}=\mu_{2}=\mu)}$
\begin{equation}
\Delta_{c}\approx \sqrt{\frac{8}{3}(e^{2}_{1}+e^{2}_{2})+9\mu^{\frac{2}{3}}}
\end{equation}
(iii) equal mass planets, equal but large eccentricities ${e}$
\begin{equation}
\Delta_{c}\approx 0.3e,
\end{equation}
where ${\mu_{1} \equiv m_{1}/m_{3}}$ and ${\mu_{2} \equiv
m_{2}/m_{3}}$ and ${e_{1}}$ and ${e_{2}}$ are the eccentricities
of the inner and outer orbit respectively.

Veras and Armitage (2004), generalising Gladman's result, derived
a criterion for two equal mass planets on initially circular
inclined orbits to achieve Hill stability.  They found that the
planets were Hill stable if their initial separation was greater
than

\begin{eqnarray}
\Delta_{crit} & = & \epsilon+\eta\sqrt{(4+\frac{\cos^{2}{I}}{2})(\epsilon+\chi\eta\mu^{\frac{2}{3}})}+
[\chi\eta\mu^{\frac{2}{3}} -\nonumber\\
& & -3\eta^{2}\mu\sqrt{\frac{4+\frac{\cos^{2}{I}}{2}}{\epsilon+\chi\eta\mu^{\frac{2}{3}}}}]+...,
\end{eqnarray}
where
\begin{displaymath}
\epsilon \equiv 2+\cos^{2}{I}-\cos{I}\sqrt{8+\cos^{2}{I}},\hspace{1cm}
\eta \equiv 1-\frac{\cos{I}}{\sqrt{8+\cos^{2}{I}}},
\end{displaymath}
\begin{displaymath}
\chi \equiv 3 \cdot 2^{\frac{2}{3}} \cdot 3^{\frac{1}{3}},
\hspace{1cm} \mu \equiv m/m_{3},
\end{displaymath}
${m_{3}}$ is the mass of the star, ${m}$ is the mass of the planets and
${I}$ the inclination of the orbits.

Finally, in a series of papers, Donnison (1984a, 1984b, 2006) made
use of the ${c^{2}H}$ criterion to determine the stability of triple
systems, where the outer body moved on a parabolic or hyperbolic
orbit with respect to the centre of mass of the other two bodies.
The first two papers dealt with coplanar systems, while the
latest one examined systems with inclined orbits. In each paper,
there was discussion about some special cases (equal masses and
large ${m_{1}}$ in paper I, equal and unequal binary masses in
paper II, equal masses, unequal binary masses, ${m_{1}}$ large in
paper III; in all cases ${m_{1}}$ belonged to the inner binary).

The main disadvantage of the ${c^{2}H}$ criterion is that it is a sufficient
but not a necessary condition for stability.  Exchange might not occur even
when the condition is violated but it certainly  cannot occur when the
condition is satisfied.  The lobes could also be open to infinity, but the
bodies may or may not escape to infinity.  Finally, things are not clear again
when the third body is started outside (inside) the lobes, since the criterion cannot
give any information whether
the third body will escape or not from the system (will keep
orbiting the binary or form a binary with one of the other masses).

The situation where one member of a triple system escapes to infinity was
investigated by several authors. They derived sufficient conditions for the
motion to be of hyperbolic-elliptic type, i.e. conditions for the distance
between one body and the centre of mass of the two other bodies to
increase indefinitely as time goes to infinity, while the distance between
the other two bodies remains bounded.  Such conditions can be found in Standish
(1971), Yoshida (1972), Griffith and North (1973), Marchal (1974).
Yoshida (1974) derived
another criterion for hyperbolic-elliptic motion under the condition that
the magnitude of the angular momentum of the three body system was above a
certain level and Bozis (1981), in a paper closely related to the
one of Yoshida (1974), he considered conditions for the smallest mass of
a triple system to escape to infinity.  Finally, a stronger escape criterion has been
proposed by Marchal and his collaborators (Marchal et al. 1984a, 1984b).
References to criteria before 1970, can be found in the above mentioned papers.

Usually, those criteria required that the distance ${\rho_{0}}$ and radial velocity ${\dot{\rho_{0}}}$
of the potential escaper
(with respect to the barycentre of the binary formed by the other two bodies)  were above
certain values at some time ${t_{0}}$.  However, for large distances ${\rho}$, there is little
difference between the criteria (Anosova 1986).

It should also be added here, that, in addition to the sufficient conditions
for escape of one body, some of the above mentioned authors also gave sufficient
conditions for ejection without escape; in such a situation, the ejected mass
reaches a bounded distance
and falls back toward the other two masses. Such conditions can be found in
Standish (1972), Griffith and North (1973) and Marchal (1974).

\subsection{Numerical integration criteria}

The numerical work involves a wide range of simulations of triple
systems.  Several authors set up numerical experiments and investigated the orbital
evolution of hierarchical triple systems.

Harrington (1972, 1975, 1977), in a series of papers, carried out numerical integrations of hierarchical
triple systems with
stellar and planetary mass ratios.  In his first paper, he integrated equal mass systems with different initial
conditions in order to determine their stability.  He considered a system to be stable if there had been no
change in the orbital elements during the period of integration, particularly in the semi-major axes or
the eccentricities.
  The following situations were also defined as unstable: escape of one body, collision, i.e. two components
got sufficiently close that it could be assumed that there were
tidal or material interactions between the bodies involved, change
to which bodies comprise the inner binary. A total of 420 orbits
were integrated for 10 to 20 revolutions of the outer orbit.  It
was found that stability was insensitive to the eccentricity of
the inner binary, for moderate eccentricity, to the argument of
periastron of either orbit and to the mutual inclination of the
two orbits (except when the inclination was within a few degrees
of a perpendicular configuration).  As a measure of stability, he
used the quantity ${q_{2}/a_{1}}$ (${q_{2}}$ was the outer
periastron distance and ${a_{1}}$  the inner semi-major axis) and
he found that stability existed above ${q_{2}/a_{1}=3.5}$ for
prograde and ${q_{2}/a_{1}=2.75}$ for retrograde orbits. In his
second paper, Harrington integrated coplanar systems with unequal
masses (with the largest mass ratio never exceeding ${100:1}$) and
based on his numerical results, he derived the following limiting
condition for stability:
\begin{equation}
\label{harr1}
q/a=[(q/a)_{0}/\log{1.5}]\log{[1+m_{3}/(m_{1}+m_{2})]},
\end{equation}
where ${q}$ is the outer periastron distance, ${a}$ is the inner
semi major axis and ${(q/a)_{0}}$  is the parameter limit for
equal masses.  The above condition was improved in the last of the
three papers, in which Harrington performed numerical simulations
for systems which consisted of a stellar binary and a body of
planetary mass (equation \ref{harr1} does not apply in this case).
The new empirical condition for stability was (regardless of which
of the components the planet was):
\begin{equation}
\label{harr2}
q_{2}/a_{1}\geq A[1+B\log{\frac{1+m_{3}/(m_{1}+m_{2})}{3/2}}]+K.
\end{equation}
 ${A}$ and ${B}$ were determined empirically, with ${A}$ being the limit on ${q_{2}/a_{1}}$ for the equal mass case
and it was taken directly from the results of the first paper and
${B}$ was then determined by a least square fit to the unequal
mass cases; ${K}$ is ${0}$ if this is to be a mean fit and is
approximately ${2}$ if it is to be an upper limit. For coplanar
prograde orbits, ${A=3.50}$ and ${B=0.70}$ and for retrograde,
${A=2.75}$ and ${B=0.64}$.  Harrington also found that retrograde
orbits were more stable than the prograde ones, a result which is
in contrast with  Szebehely's and Zare's predictions, as they
found that prograde orbits were more stable than retrograde
orbits. However, the results for equal masses and direct orbits
were in good agreement, although Szebehely's results allow a
slightly closer outer orbit.  Of course, it should be borne in
mind that the ${c^{2}H}$ criterion is a sufficient stability
condition, based on the possibility of exchange of bodies.  It
should also be pointed out that the definition of stability given
by Harrington is a bit ambiguous.  He classifies a triple system
as stable if there is no ``significant change'' in the orbital
elements during the period of integration, and particularly in the
semi-major axes and eccentricities.  Another point that raises
some concern is that the integrations were performed for only 10
or 20 outer orbital periods.  This could prove inadequate,
although Harrington suggested that instabilities of this kind
(exchange etc.) set in very quickly.

Graziani and Black (1981), in the context of planet formation and
extrasolar planets, used numerical integrations to model planetary
systems (star and two planets, which had the same mass in most of
the numerical simulations) with prograde, coplanar and initially
circular orbits. The systems were integrated for at least 100
revolutions of the longest period planet, or until instability was
evident. The authors  classified a system as unstable if there was
clear evidence for secular  changes in any of the orbits during
the numerical integration. Based on their results, they obtained
the following condition for stability:
\begin{equation}
\label{black1}
\mu=0.5\frac{m_{1}+m_{2}}{M_{*}}<\mu_{crit}=0.175\Delta^{3}(2-\Delta)^
{-\frac{3}{2}},\hspace{0.3 cm} \mu\leq 1
\end{equation}
where the planets ${m_{1}}$ and ${m_{2}}$ orbit the star ${M_{*}}$.  The
parameter ${\Delta}$ gives the minimum initial separation between the companions
in units of their mean distance  from the central star, while ${\mu}$ is the
mean mass of the two companions in units of the mass of the star.  More specifically,
\begin{displaymath}
\Delta=2\frac{R-1}{R+1},\hspace{0.5 cm} R=\frac{R_{2}}{R_{1}}
\end{displaymath}
with ${R_{1}}$ and ${R_{2}}$  being the semi-major axes of the
inner and outer orbit respectively.  Systems with ${\mu \geq
\mu_{crit}}$ became unstable within a few tens of planetary
orbits. Black (1982) modified the above condition to apply for
${\mu\geq 1}$.  The modified stability condition is:
\begin{equation}
\label{black2}
\mu \leq \mu_{crit}=0.083\frac{\Delta^{3}}{(2-\Delta)^{3}}.
\end{equation}
Both the above stability conditions were confirmed by more
integrations (Pendleton and Black 1983).  However, equations (\ref{black1}) and (\ref{black2}) were in disagreement
with equation (\ref{harr2}), except a narrow range around ${\mu=1}$.

Donnison and Mikulskis (1992) produced a modified version of
equations (\ref{black1}) and (\ref{black2}), based on numerical
integrations of circular, coplanar and prograde systems.  A system was considered to be
unstable when there was a change of more than ${10\%}$ in either of the semi-major axes
or/and either of the eccentricities altered
by more than 0.1.  Each numerical model was integrated for at
least 1000 inner binary orbits or until the existence of
instability was evident (which usually happened within the first
100 orbits). They derived the following values for ${\mu_{crit}}$:
\begin{equation}
\label{dm1}
\mu_{crit}=0.479\frac{\Delta^{3}}{(2-\Delta)^{\frac{3}{2}}},\hspace{0.5 cm} \mu\leq 1
\end{equation}
and
\begin{equation}
\label{dm2}
\mu_{crit}=0.364\frac{\Delta^{3}}{(2-\Delta)^{3}},\hspace{0.5 cm} \mu\geq 1.
\end{equation}
Donnison and Mikulskis (1994), following the same procedure as above, produced the following formulae
for ${\mu_{crit}}$ in the case of retrograde orbits:
\begin{equation}
\label{dm3}
\mu_{crit}=0.7692\frac{\Delta^{3}}{(2-\Delta)^{\frac{3}{2}}},\hspace{0.5 cm} \mu\leq 1
\end{equation}
and
\begin{equation}
\label{dm4}
\mu_{crit}=0.6848\frac{\Delta^{3}}{(2-\Delta)^{3}},\hspace{0.5 cm} \mu\geq 1.
\end{equation}
The results of Donnison and Mikulskis (1992, 1994) were in good
agreement with the results of Black and his collaborators (for
prograde orbits of course), but quite different from Harrington's
results, except  in the equal mass case.  There was also agreement
with the theory of Szebehely and Zare (1977), but only for
prograde orbits.

Dvorak  (1986) investigated the stability of P-type orbits in
stellar binary systems, i.e. planet orbiting the binary system, in
the context of the elliptic restricted three body problem. He
performed numerical integrations of  planets on initially circular
orbits orbiting an equal mass binary system. The integration time
span was 500 binary periods and a planetary orbit was classified
as stable if its eccentricity remained smaller than ${0.3}$
throughout the whole integration time.  His results showed a
region of stability far away from the primaries, a region of
instability closer to the primaries and a  chaotic (in the sense
of unpredictability) zone between those two regions. This chaotic
zone was limited by the lower critical orbit (LCO), defined as the
largest unstable orbit for all starting positions of the planet,
and the upper critical orbit (UCO), defined as the orbit with the
smallest semimajor axis for which the system was stable for all
starting positions.  A least squares parabolic fit to the
numerical integration results yielded:
\begin{eqnarray}
LCO & = & (2.09\pm 0.30)+(2.79\pm 0.53) e-(2.08\pm 0.56) e^{2}\\
UCO & = & (2.37\pm 0.23)+(2.76\pm 0.40) e-(1.04\pm 0.43) e^{2},
\end{eqnarray}
where ${e}$ is the eccentricity of the primaries and the distance
is given in AU. Each coefficient is listed along with its formal
uncertainty. Although the above formulae were derived for systems
where the primaries had equal masses, additional numerical
integrations of P-type orbits in systems  with unequal mass
primaries (Dvorak et al. 1989) showed no dependence of the
critical orbits on the mass ratio of the primaries.  Finally,
concerning P-type orbits, Pilat-Lohinger et al. (2003)
investigated the stability of such orbits in three dimensional
space.  They integrated initially circular planetary orbits in
equal mass binary systems, with a binary eccentricity varying from
0 to 0.5. The mutual inclination of the orbits was in the range
${0^{\circ}-50^{\circ}}$.  The orbits were classified as in Dvorak
(1986), i.e. stable, chaotic and unstable, where stable meant that
the planet did not suffer from a close encounter with one of the
primaries for the whole integration time span (50000 periods of
the primaries).   It turned out that the inclination did not
affect the stability limit significantly.

Rabl and Dvorak (1988), by using numerical integrations,
established stability zones for S-type orbits in stellar binary
systems (planet orbiting one of the stars of the binary system) .
The setup of their systems was similar to the one in Dvorak
(1986), i.e. initially circular orbit for the massless particle
and equal mass primaries.  The maximum binary eccentricity
considered was 0.6. An initially circular S-type orbit was
classified as stable, if it remained elliptical with respect to its
mother primary during the whole integration time of 300 periods of
the primary bodies. Based on their results, they derived the
following formulae:
\begin{eqnarray}
LCO & = & (0.262\pm 0.006)-(0.254\pm 0.017) e-(0.060\pm 0.027) e^{2}\\
UCO & = & (0.336\pm 0.020)-(0.332\pm 0.051)e-(0.082\pm 0.082)e^{2},
\end{eqnarray}
where ${e}$ is the eccentricity of the stellar binary. Note that
the meaning of LCO and UCO is different compared to the P-type
orbit case (the stable orbits lie inside LCO, while the unstable
ones outside UCO).   As in Dvorak (1986), the results showed the
existence of a grey (chaotic) area between LCO and UCO.
Pilat-Lohinger and Dvorak (2002) performed more numerical
experiments on S-type orbits.  Their models took into
consideration varying binary mass ratios (${0.1-0.9}$) and,
besides a varying primary eccentricity, the planetary mass had an
eccentricity from 0 to 0.5.  The integration time was 1000 binary
periods.  They found that an increase in the eccentricities
reduced the stability zone (the planetary eccentricity had less
influence than the binary eccentricity, but it reduced the
stability zone in a similar way). The results were also in
agreement with the results of Rabl and Dvorak (1988).  However, a
quick inspection of the result tables in  Pilat-Lohinger and
Dvorak (2002), may suggest that the primary mass ratio has an
effect on the stability zones, in contrast to what was mentioned
above in the case of P-type orbits.

Holman and Wiegert (1999), also investigated  the stability of P-type and S-type orbits in stellar binary systems.
They performed numerical simulations of  particles  on initially circular and prograde orbits around the binary or
around one of the stars, in
the binary plane of motion and with different initial orbital longitudes.  The binary mass ratio was taken in the
range ${0.1 \leq \mu \leq 0.9}$ and the binary eccentricity in the range ${0.0 \leq e \leq 0.7-0.8}$.  The integrations
lasted for ${10^{4}}$ binary periods.  If a particle survived the
whole integration time at all initial longitudes, then the system was classified as stable.
  Using a least squares fit to their data, they obtained:
(i) for the inner region (S-type orbit):
\begin{eqnarray}
a_{c} & = & [(0.464\pm 0.006)+(-0.380\pm 0.010)\mu+(-0.631\pm 0.034)e+\nonumber\\
& & +(0.586\pm 0.061)\mu e+(0.150\pm 0.041)e^{2}+\nonumber\\
& & +(-0.198\pm 0.074)\mu e^{2}]a_{b}
\label{hol1}
\end{eqnarray}
(ii) for the outer region (P-type orbit):
\begin{eqnarray}
a_{c} & = & [(1.60\pm 0.04)+(5.10\pm 0.05)e+(-2.22\pm 0.11)e^{2}+\nonumber\\
& & +(4.12\pm 0.09)\mu +(-4.27\pm 0.17)e\mu+(-5.09\pm 0.11)\mu^{2}+\nonumber\\
& & +(4.61\pm 0.36)e^{2}\mu^{2}]a_{b},
\label{hol2}
\end{eqnarray}
where ${a_{c}}$ is the critical semi-major axis, ${a_{b}}$ is the
binary semi-major axis, ${e}$ is the binary eccentricity and
${\mu=m_{2}/(m_{1}+m_{2})}$.  Equation (\ref{hol1}) is valid to
${4 \%}$ typically and to ${11 \%}$ in the worst case over the
range  of ${0.1\leq \mu \leq 0.9}$ and ${0.0\leq e \leq 0.8}$,
while equation (\ref{hol2}) is valid to ${3 \%}$ typically and to
${6 \%}$ in the worst case over the same ranges.  An interesting
finding was that, in the outer region, `islands' of instability
existed outside the inner stable region; this phenomenon was
attributed to mean motion resonances and indicated that there was
not a sharp boundary between stable and unstable regions.  It
should be mentioned here that equation (\ref{hol2}), as presented
in the paper of Holman and Wiegert, appears not to depend on
${a_{b}}$ at all.  However, this is probably a misprint, as
equation (\ref{hol1}) might suggest.  The results of Holman and
Wiegert are in good agreement with the results of Dvorak (1986)
and Rabl and Dvorak (1988).  Figures 1 demonstrate that agreement.

\begin{figure}
\begin{center}
\includegraphics[width=80mm,height=60mm]{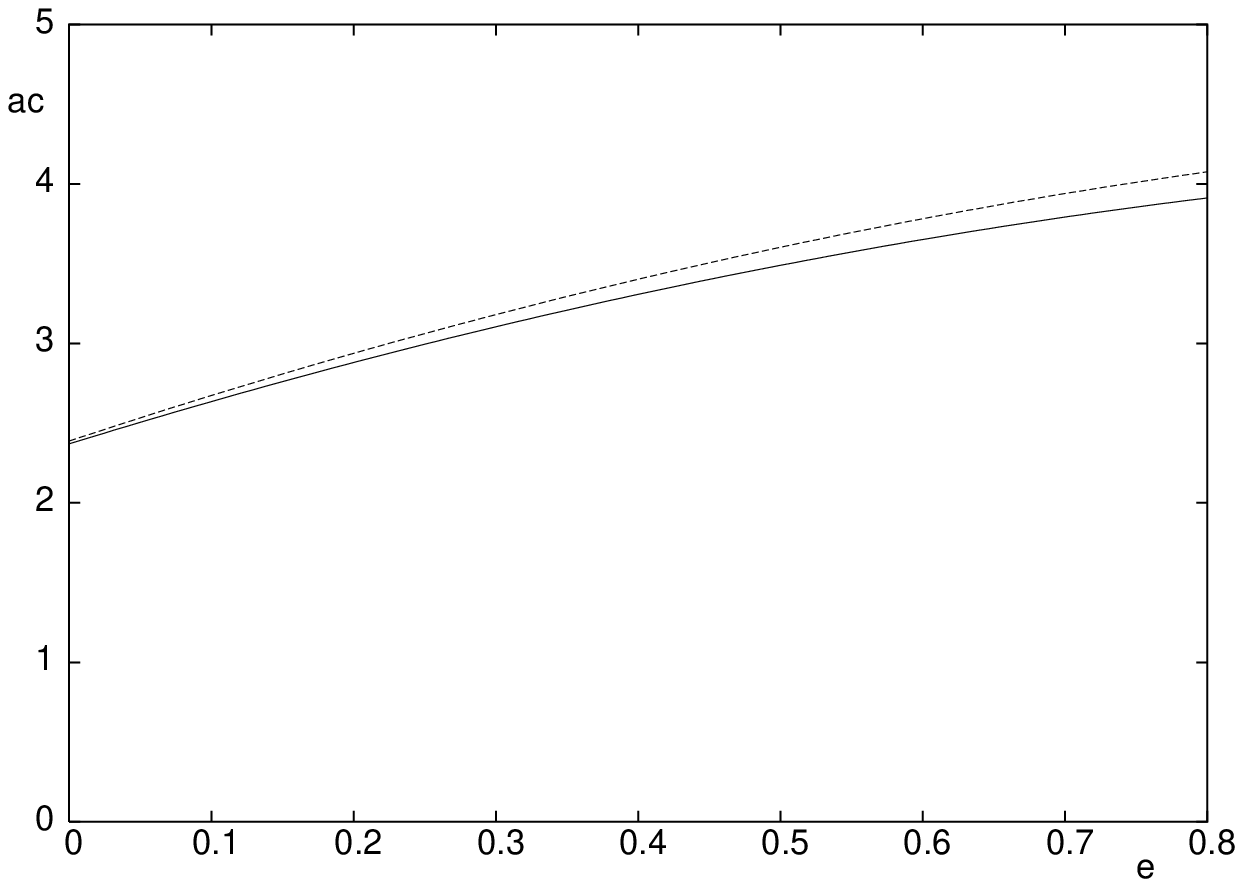}
\includegraphics[width=80mm,height=60mm]{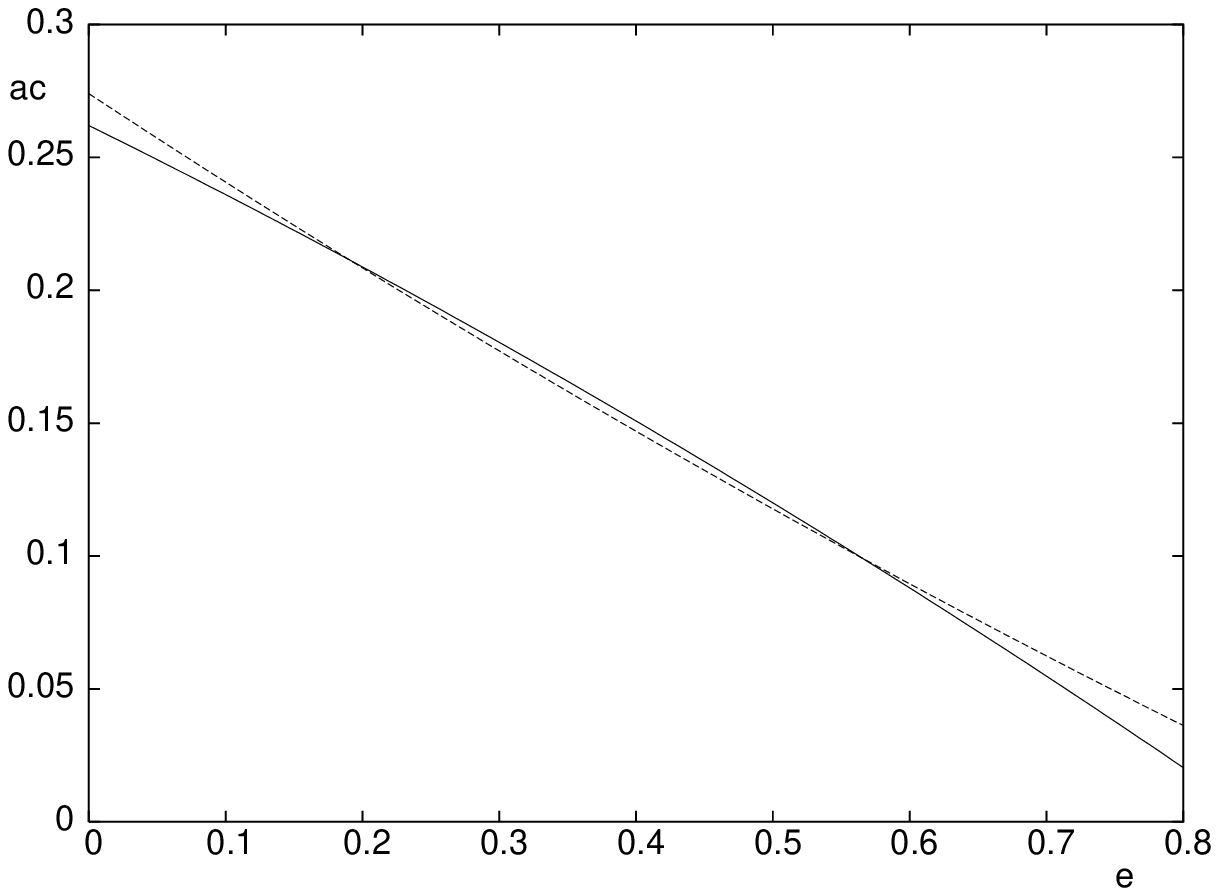}
\caption[]{Critical semi-major axis ${a_{c}}$ against binary
eccentricity ${e}$ for a particle orbiting the binary.  The top
graph is for P-type orbits and the bottom one is for S-type
orbits. The continuous lines comes from the results obtained from
Dvorak (1986) and Rabl and Dvorak (1988), while  the
Holman-Wiegert
 results are shown with the dashed lines.  For both graphs,
${\mu=0.5}$ and the binary semi-major axis is 1 AU.}
\end{center}
\end{figure}

Kiseleva and her collaborators, performed numerical integrations of hierarchical
triple systems with coplanar, prograde and initially circular orbits (Kiseleva et al.
 1994a, 1994b).  The mass ratios were within the range ${1:1-100:1}$.  A system was classified
as stable if it preserved its initial hierarchical configuration during the whole of the
integration time span, which was normally 100 outer binary orbital periods, but certain
cases were followed for 1000 or even for 10000 outer orbits (however, it appeared that
the longer integration time had little effect on the stability boundary).
These numerical calculations
were later extended to eccentric binaries, inclined orbits (from ${0^{\circ}}$ to
${180^{\circ}}$) and different initial phases, and an empirical condition for stability
 was derived (Eggleton and Kiseleva 1995):

\begin{equation}
\label{egg1}
Y_{0}^{min} \approx 1+\frac{3.7}{q_{out}^{1/3}}-\frac{2.2}{1+q_{out}^{1/3}}
+\frac{1.4}{q_{in}^{1/3}}\frac{q_{out}^{1/3}-1}{q_{out}^{1/3}+1},
\label{kegg}
\end{equation}
where
${Y_{0}^{min}}$ is the critical initial ratio of the periastron distance
of the outer orbit to the apastron distance of the inner orbit,
\begin{displaymath}
q_{in}=\frac{m_{1}}{m_{2}} \geq 1,\hspace{0.5 cm} q_{out}=\frac{m_{1}+m_{2}}
{m_{3}}.
\end{displaymath}
  ${Y_{0}^{min}}$ is related to the critical initial period ratio ${X_{0}^{min}}$ by the
following relation:
\begin{equation}
(X_{0}^{min})^{\frac{2}{3}}=(\frac{q_{out}}{1+q_{out}})^{\frac{1}{3}}
\frac{1+e_{in}}{1-e_{out}}Y_{0}^{min},
\end{equation}
where ${e_{in}}$ and ${e_{out}}$ are the eccentricities of the inner and outer orbit respectively.
The coefficients of equation (\ref{egg1}) were obtained rather empirically
, based on the numerical results that the authors had at their disposal.  As for the
 effect of certain characteristics on the stability boundary, such as the orbital
eccentricities, it was determined by the examination of a small
number of mass ratios that the authors believed to be reasonably
representative. The criterion  appears to be reliable to about
${20 \%}$ for a wide range of
 circumstances, which is not very bad,
considering the amount of parameters and the complex nature of the critical
surface.  It probably does not work very well in situations where there is a resonance or
commensurability, but these are more common in systems with extreme mass ratios
(e.g. star and planets), while the intention of the authors (as stated in their paper) was
to investigate triple systems of comparable masses.
It should be pointed out here that there is
a misprint in formula (\ref{egg1}) as given in Eggleton and Kiseleva (1995): the sign
of the term ${2.2/(1+q_{out}^{1/3})}$ is plus, while it should
be minus (Aarseth 2003).

\subsection{Chaotic Criteria}

In the two previous sections, we presented stability criteria that were derived either analytically  or based
on results from numerical simulations.  In this section, we discuss criteria that are based on the concept of chaos.

Wisdom (1980), applied the Chirikov resonance overlap criterion for the onset of stochastic behaviour (Chirikov 1979)
to the planar circular restricted three body problem.  He derived the following estimate of when resonances should
start to overlap (the derivation holds for small eccentricities ${e\leq 0.15}$):
\begin{equation}
s_{overlap} \simeq 0.51\mu^{-2/7},
\end{equation}
where ${\mu=m_{2}/(m_{1}+m_{2})<<1}$.  By using Kepler's third law, this can be expressed in terms of
the semi-major axis separation as (Murray and Dermott 1999)
\begin{equation}
\label{wis}
\Delta a_{overlap} \simeq 1.3 \mu^{2/7}a_{2},
\end{equation}
where ${a_{2}}$ is the semi-major axis of the perturber.  Hence,  when the particle is in the region
${a_{2}\pm \Delta a_{overlap}}$, the orbit is chaotic.
A similar result to the one of Wisdom, was obtained through the use of a mapping, which was based on the
approximation that perturbations to the massless body are localised near conjunction with the perturber
(Duncan et al. 1989).  It was found that
\begin{equation}
\Delta a_{overlap} \simeq 1.24 \mu^{2/7}a_{2},
\end {equation}
which is in agreement with equation (\ref{wis}).

Mardling and Aarseth (1999) approached the stability problem in a
different way, by noticing that stability against escape in the
three body problem is analogous to stability against chaotic
energy exchange in the binary-tides problem.  The way energy and
angular momentum are exchanged between the two orbits of a stable
(unstable) hierarchical triple system is similar to the way they
are exchanged in a binary undergoing normal (chaotic) tide-orbit
interaction.  Having that in mind, they derived the following
semi-analytical formula for the critical value of the outer
pericentre distance ${R_{p}^{crit}}$:
\begin{equation}
\label{mard}
R_{p}^{crit}=C\left[(1+q_{out})\frac{1+e_{out}}{(1-e_{out})^
{\frac{1}{2}}}\right]^{\frac{2}{5}}
\end{equation}
where  ${q_{out}=m_{3}/(m_{1}+m_{2})}$ is the mass ratio of the
outer binary and ${e_{out}}$ is the outer binary eccentricity. If
${R_{p}^{crit}\leq R_{p}^{out}}$, then the system is considered to
be stable.  The above formula is valid for prograde and coplanar
systems and it applies to escape of the outer body. C was
determined empirically and it was found to be 2.8. A small
heuristic correction of up to ${30\%}$ was then applied for
non-inclined orbits, to account for the increased stability
(Aarseth and Mardling 2001, Aarseth 2004).  Also, as stated in
Aarseth and Mardling (2001), the criterion ignores  a weak
dependence on the inner eccentricity  and inner mass ratio.
Finally, we should mention here, that, numerical tests have showed
that the criterion is working well for a wide range of parameters,
but it has not been tested for systems with planetary masses so
far (Aarseth 2004), probably because the authors were mainly
interested in using the formula in star cluster simulations.

We would like to mention here that, Mardling (2007) has derived a
resonance overlap criterion for the general three body problem.

We should point out, that the presence of chaos does not
necessarily indicate instability, e.g. see Murray (1992),
Gladman(1993).  The reader should also recall the results of
Dvorak (1986) and Rabl and Dvorak (1988), with the zones of
unpredictability between the stable and unstable orbits.  However,
that kind of behaviour appears to depend on various parameters of
the system, such as the mass ratios of the system.  For example,
Mudryk and Wu (2006), in their study of a planet orbiting one of
the components of a stellar binary system, found little evidence
of 'bound chaos' near the instability boundary (except in the case
where the perturber is very small compared to the star, i.e. the
case discussed by Gladman or covered by Wisdom's criterion) and as
a result of that, they adopted the boundary of resonance overlap
as the boundary of instability.  That appears to be the case with
Mardling and Aarseth too. A nice discussion in resonances and
instability can be found in Mardling (2001).

\section{Summary}

We have attempted to collect and present the various criteria that
have been derived for the stability of hierarchical triple systems
over the past few decades.

Tables 1, 2 and 3 present the various criteria in a rather concise
manner. Each Table consists of four columns, i.e. the 'Name'
column, which gives the name of the relative paper(s), the
'Model/Restrictions' column, which gives a brief description of
the systems for which the criterion is applicable (a blank line
indicates that the criterion applies to the general case, without
any restrictions), the 'Stability Type' column, which states what
stability means for a specific criterion and finally the
'Comments' column, where we give any extra information we consider
important.

Table 1 lists the criteria that were derived analytically. Most of
them were based on a generalisation of the concept of zero
velocity surfaces of the circular restricted three body problem,
 with the quantity ${c^{2}H}$ playing the role of the Jacobi constant.  As stated in the
corresponding section, the ${c^{2}H}$ criterion is a sufficient
condition and therefore, no conclusion can be drawn when it is
violated.  The Marchal and Bozis (1982) criterion is a good choice
for one who intends to use a criterion from that specific
category.  However, depending on the system investigated, the
other criteria could also be a useful alternative and even easier
to apply. Table 1 also lists sufficient criteria for escape of one
of the bodies.  Although those criteria are not very useful on
their own, because of their nature (they require some conditions
to be satisfied at a moment ${t_{0}}$), they could be used as part
of a computer code (e.g. for cluster simulations); however, their
sufficient nature is a major disadvantage for that type of use.

Table 2 presents criteria that were based on results from
numerical integrations.  A task that is not particularly easy, as
a triple system has many parameters to be taken into consideration
(mass ratios and orbital parameters) and covering the whole of the
parameter space at once is a rather difficult thing.  Sometimes
the various criteria were in agreement with each other, sometimes
they were not.  This can be attributed to many factors.  The main
one, in our opinion, is the different meaning that stability may
have for different people.  Szebehely (1984) gave 47 different
definitions for stability in his 'Dictionary of Stability'.  As
the reader has probably noticed, almost each author mentioned in
section (2.2), gave a different definition of what he considered
as stable system.  Another issue that raises concern is the
integration time span. A system may appear to be stable for a
certain time span, but becomes unstable when the integration is
extended over longer timescales.  Also, the choice of initial
conditions may have an effect on the outcome.  Finally, as stated
in Kiseleva et al. (1994a), a matter of concern about those
criteria is the fact that they involve instantaneous and not mean
orbital parameters.  The last two criteria of the table are
probably the best from the numerical ones, the Eggleton-Kiseleva
for stellar systems and the Holman-Wiegert for planets in binary
systems (keep in mind that the planets are on intially circular
orbits).

We would like to open a parenthesis here and mention that the
stability of planets in binary systems is an area of research that
is expected to become more and more important in the future, as
there is an increasing number of exoplanets that are members of
binary or multiple stellar systems (e.g. see Eggenberger et al.
2004). It appears that none of the above mentioned stability
criteria, analytical or numerical, can cover the issue on its own.
For instance, the planetary eccentricity is an important parameter
not appearing in the criteria, although many exoplanets have
eccentric orbits (of course most of the criteria were developed
when none or very few exoplanets had been discovered by that
time). Therefore, at the moment, one should choose the criterion
(or a combination of different criteria) that fits the system he
investigates better.

\begin{table}
\caption[]{Analytical criteria overview} \vspace{0.1 cm}
{\footnotesize
\begin{tabular}{l l l l}\hline
 Name & Model/Restrictions & Stability & Comments \\
      &  & Type &  \\
\hline
Szebehely and &  coplanar orbits & Hill & 2b approx. for energy \\
Zare 1977 &                  &      & and ang. momentum\\
\hline
Marchal and   &  & Hill & limit cases discussed \\
Bozis 1982 &                  &      & \\
\hline
Walker & coplanar, corotational & Hill & series approx. for the \\
et al. 1980 & init. circular orbits & & controlling parameter \\
\hline
Roy et al. 1984 & coplanar, corotational & Hill & in agreement with \\
  &  orbits  & & the ${c^{2}H}$ criterion \\
\hline
Szebehely 1978 & circular restricted 3bp & Hill &  \\
 & satellite-planet + star  &      & \\
\hline
Markellos and & circular restricted 3bp & Hill & more accurate \\
Roy 1981 & satellite-planet + star  &      & result than Szeb. 1978\\
\hline
Walker 1983 & coplanar,  & Hill & in agreement with \\
 & init. circular orbits & & Markellos and Roy \\
 &  ${m_{3}>>m_{1}+m_{2}}$ &      & \\
\hline
Donnison 1988 & coplanar,  & Hill & in agreement with \\
 & init. circular orbits & & the previous three papers \\
 &  ${m_{3}>>m_{1}+m_{2}}$   &      & \\
\hline
Donnison and & coplanar orbits & Hill & 2b approx. for energy \\
Williams & ${m_{1}>>m_{2}, m_{3}}$  &   & and ang. momentum \\
1983, 1985  &  &   &  \\
\hline
Gladman 1993 & star + two planets & Hill & based on Marchal   \\
            & init. circular orbits, &   & and Bozis 1982 \\
  & equal planetary masses &   &  \\
  & and small ${e}$,  &   &              \\
  & equal planetary masses &   &  \\
  & and equal but large ${e}$ &   &  \\
\hline
Veras and & star + two equal mass & Hill &  generalisation of  \\
Armitage 2004  & planets &   & Gladman's result \\
      & initially circular &   &  \\
  & and inclined orbits &   &  \\
\hline
Donnison 1984a & coplanar, non-  & Hill &  \\
 & closed outer orbit & &   \\
 & equal masses & &   \\
 & large ${m_{1}}$ & &   \\
\hline
Donnison 1984b & coplanar, non-  & Hill &  \\
 & closed outer orbit & &   \\
 & equal masses & &   \\
 & equal binary masses & &   \\
 & unequal binary masses & &   \\
\hline
Donnison 2006 & non-coplanar  & Hill &  \\
 & parabolic outer orbit & &   \\
 & equal masses & &   \\
& unequal binary masses & &   \\
 & large ${m_{1}}$ & &   \\
\hline
Standish 1971 &  & escape &  \\
\hline
Yoshida 1972 &  & escape &  \\
\hline
Griffith and &  & escape &  \\
North 1973 &  &  &  \\
\hline
Marchal 1974 &  & escape &  \\
\hline
Yoshida 1974 &  & escape &  \\
\hline
Bozis 1981 & ${m_{1} \geq m_{2} \geq m_{3}}$ & escape &  \\
           &                                 & of ${m_{3}}$ & \\
\hline
Marchal et al. &  & escape & stronger than the \\
1984a, 1984b &  &  & previous relevant \\
             &  &  & criteria \\
\hline

\end{tabular}}
\end{table}

\begin{table}
\caption[]{Numerical integration criteria overview} \vspace{0.1
cm} {\footnotesize
\begin{tabular}{l l l l}\hline
 Name & Model/Restrictions & Stability & Comments \\
      &  & Type &  \\
\hline
Harrington &  & no significant  & inclination not  \\
 1977&    & change in ${a,e}$,   & important,\\
 &    & no escape & ${T_{int}=10-20}$\\
 &    & no collision   & outer orbital \\
 &    & no change in   & periods\\
 &    & hierarchy  & \\
\hline
Graziani and & Star + two planets, & Laplace  & ${T_{int}=}$ at least  \\
Black 1981   & prograde,           &          & 100 outer periods\\
(GB 1981)    & init. circular,     &          & \\
             & coplanar orbits     &          & \\
\hline
Black 1982 & ${m_{3}>m_{1},m_{2}}$, & Laplace  & non-numerical,  \\
           & prograde,           &          & extends GB 1981\\
             & init. circular,   &          & \\
            & coplanar orbits    &          & \\
\hline
Donnison and    & same as GB 1981 & change less than         & ${T_{int}=}$ at least  \\
Mikulskis 1992, & plus retrograde & ${10 \%}$ in a and       & ${10^{3}}$ inner periods \\
1994            & orbits          &  less than 0.1 in ${e}$  & or until \\
                &                 &                          & instability \\
                &                 &                          & evident \\
\hline
Dvorak 1986 & elliptic restricted &   ${e_{planet}<0.3}$  &  ${T_{int}=500}$  \\
(DV 1986)   & P-type in equal     &                       & bin. periods\\
            & mass stellar        &                       & \\
            & binaries,           &                       & \\
            & coplanar, init.     &                       & \\
            & circular plan.      &                       & \\
            & orbit               &                       & \\
\hline
Rabl and    & same as DV 1986, & planet elliptic &  ${T_{int}=300}$  \\
Dvorak 1988 & but for S-type   & with respect    & bin. periods \\
            & orbits           & to mother prim. &              \\
            &                  &                 &              \\
\hline
Holman and   & elliptic restricted & planet survives      & ${T_{int}=10^{4}}$  \\
Wiegert 1999 & P,S-type in         & at all init. longit. & bin. periods\\
             &  stellar bin.       & for ${T_{int}}$       & \\
             & coplanar, init.     &                      & \\
             & circular plan.      &                      & \\
             & orbit               &                      & \\
\hline
Eggleton and   & stellar mass   & change in & mostly ${T_{int}=}$  \\
Kiseleva 1995  & ratios         & hierarchy & 100 outer periods \\
\hline
\end{tabular}}
\end{table}

\begin{table}
\caption[]{Chaotic criteria overview} \vspace{0.1 cm}
{\footnotesize
\begin{tabular}{l l l l}\hline
 Name & Model/Restrictions & Stability & Comments \\
      &  & Type &  \\
\hline
Wisdom 1980 & planar circular         &  res. overlap &   \\
            & restricted,             &               & \\
            & ${e_{part.} \leq 0.15}$ &               & \\
\hline
Duncan et & same as     &  res. overlap & agrees with  \\
al. 1989  & Wisdom 1980 &               & Wisdom 1980\\
\hline
Mardling and  & stellar &  res. overlap       & semi-analytical  \\
Aarseth 1999  & systems, &  and escape        & criterion \\
              & copl. orbits &          &  \\
\hline
\end{tabular}}
\end{table}

Finally, Table 3 lists criteria that involve the concept of chaos.
In that context, instability in a three body system was thought to
be the consequence of the overlap of sub-resonances within mean
motion resonances.  It was also mentioned that the presence of
chaos in some cases, would not necessarily indicate instability.

We hope that this work can serve as a useful guide for anyone
interested in the issue of the stability of hierarchical triple
systems.

\begin{acknowledgements}

The author wants to thank the Institute for Materials and
Processes at the School of Engineering and Electronics of
Edinburgh University, where most of this work took place. The
author also thanks the anonymous referees for their useful
comments on various aspects of this work.

\end{acknowledgements}

\end{document}